\documentclass[11pt]{article}
\usepackage{amssymb}
\usepackage{colortbl}
\usepackage{amsfonts,amsmath, longtable}

\newcommand{\tr}{{\rm tr}}

\newcommand{\Mat}{ {\rm Mat}(N,\mathbb C) }

\newcommand{\mC}{\mathbb C}

\def\beq{\begin{equation}}
\def\eq{\end{equation}}
\def\p{\partial}

\begin{document}

\setcounter{page}{1}

\begin{center}

\

\vspace{-10mm}

{\Large{\bf Gauge equivalence between 1+1 rational Calogero-Moser }}

\vspace{3mm}

{\Large{\bf field theory and higher rank Landau-Lifshitz equation}}

% IRF-Vertex correspondence for 1+1 field Calogero-Moser model

 \vspace{12mm}

 {\Large {K. Atalikov}}$\,^{\diamond\,\bullet}$
\qquad\quad\quad
 {\Large {A. Zotov}}$\,^{\diamond\,\bullet\,*}$

  \vspace{5mm}

$\diamond$ -- {\em Steklov Mathematical Institute of Russian
Academy of Sciences,\\ Gubkina str. 8, 119991, Moscow, Russia}

%$*$ -- {\em Center for Advanced Studies, Skoltech, 143026, Moscow, Russia}

$\bullet$ -- {\em NRC ''Kurchatov Institute'',\\
Kurchatova sq. 1, 123182, Moscow, Russia}

$*$ -- {\em National Research University Higher School of Economics,\\ 119048, Usacheva str. 6, Moscow, Russia}

%НИЦ "Курчатовский Институт"
%123182 Россия, Москва, пл. Академика Курчатова, д. 1
%Russia, 123182, Moscow, pl. Kurchatova, 1

   \vspace{3mm}

 {\small\rm {e-mails: kantemir.atalikov@yandex.ru, zotov@mi-ras.ru}}

\end{center}

\vspace{0mm}

\begin{abstract}
In this paper we study 1+1 field generalization of the rational $N$-body Calogero-Moser model. We show that this model is gauge equivalent to some special higher rank matrix Landau-Lifshitz equation. The latter equation is described in terms of ${\rm GL}_N$ rational $R$-matrix, which turns into the 11-vertex $R$-matrix in the $N=2$ case. The rational $R$-matrix satisfies the associative Yang-Baxter equation, which underlies construction of
the Lax pair for the Zakharov-Shabat equation.  The field analogue
of the IRF-Vertex transformation is proposed. It allows to compute explicit change of variables between the field Calogero-Moser model and the Landau-Lifshitz equation.
\end{abstract}

%

%\newpage
%{\small{
%\tableofcontents
%}}

%\vspace{5mm}

%\newpage

%%%%%%%%%%%%%%%%%%%%%%%%%%%%%%%%%%%%%%%%%%%%%%%%%%%%%%%%%%%%%%%%%%%%%%%%%%%%%%%%%%%%%%%%%%%%%%%%%%%%%%
%%%%%%%%%%%%%%%%%%%%%%%%%%%%%%%%%%%%%%%%%%%%%%%%%%%%%%%%%%%%%%%%%%%%%%%%%%%%%%%%%%%%%%%%%%%%%%%%%%%%%%
\section{Calogero-Moser field theory}
\setcounter{equation}{0}

 {\bf The 1+1 field generalization\footnote{1+1 or 2d means 1 dimension for space variable and 1 dimension for time variable. In this respect mechanics is 0+1.} of the Calogero-Moser model} was proposed in \cite{Krich2, Akhmetshin}, see also \cite{LOZ}. The Hamiltonian is given by the following expression\footnote{In \cite{Krich2, Akhmetshin} the elliptic model was considered. In this paper we deal with its rational limit.}:
 \beq\label{e01}
 \begin{array}{c}
  \displaystyle{
{\mathcal H}^{\hbox{\tiny{2dCM}}}=\oint{\rm d}x\,  H^{\hbox{\tiny{2dCM}}}(x) \,,
}
 \\
   \displaystyle{
H^{\hbox{\tiny{2dCM}}}(x)= \sum_{i=1}^{N} {p}_{i}^{2}\left(c-k {q}_{i x}\right)-\frac{1}{N c}\left(\sum_{i=1}^{N} {p}_{i}\left(c-k {q}_{i x}\right)\right)^{2}-
}
 \\
   \displaystyle{
-\sum_{i=1}^{N} \frac{ k^{4} {q}_{i x x}^{2}}{4\left(c-k {q}_{i x}\right)}+\frac{k^{3}}{2} \sum_{i \neq j}^N\frac{{q}_{i x} {q}_{j x x}-{q}_{j x} {q}_{i x x}}{{q}_{i}-{q}_{j}}\,-
}
 \\
   \displaystyle{
-\frac{1}{2} \sum_{i \neq j}^N\frac{1}{\left({q}_{i}-{q}_{j}\right)^2}\left[\left(c-k {q}_{i x}\right)^{2}\left(c-k {q}_{j x}\right)+\left(c-k {q}_{i x}\right)\left(c-k {q}_{j x}\right)^{2}-c k^{2} \left({q}_{i x}- {q}_{j x}\right)^{2}\right]\,,
  }
 \end{array}
\eq
 where $x$ is the (space) field variable. It is a coordinate on a unit circle. Dynamical variables are the ($\mC$-valued) fields $p_i=p_i(x)$, $q_i=q_i(x)$,
 $i=1,...,N$, and the lower index ''$x$'' means derivative with respect to $x$. For instance, $q_{jxx}=\p_x^2q_j(x)$.
 The parameter $c\in\mC$ is a coupling constant and $k\in\mC$ is an auxiliary parameter, which can be fixed as $k=1$ but we keep it as it is. The momenta $p_i$ and coordinates $q_j$ are canonically conjugated fields:
 \beq\label{e02}
 \begin{array}{c}
  \displaystyle{
\{q_i(x),p_j(y)\}=\delta_{ij}\delta(x-y)\,,\qquad \{p_i(x),p_j(y)\}=\{q_i(x),q_j(y)\}=0\,.
  }
 \end{array}
\eq
Equations of motion (the Hamiltonian equations ${\dot f}=\{f,H\}$) take the following form:
 \beq\label{e021}
 \begin{array}{c}
  \displaystyle{
\dot{q}_{i}=2 p_{i}\left(c-k q_{i x}\right)-\frac{2}{N c} \sum_{l=1}^{N} p_{l}\left(c-k q_{l x}\right)\left(c-k q_{i x}\right)\,,
  }
 \end{array}
\eq
$$
\dot{p}_{i}=-2 k p_{i} p_{i x}+\frac{2 k}{N c}\left\{\sum_{l=1}^{N} p_{i} p_{l}\left(c-k q_{l x}\right)\right\}_{x}+k \left\{\frac{k^{3} q_{i x x x}}{2\left(c-k q_{i x}\right)}+\frac{k^{4} q_{i x x}^{2}}{4\left(c-k q_{i x}\right)^{2}}\right\}_{x}+
$$
 \beq\label{e022}
 \begin{array}{c}
  \displaystyle{
+2 \sum_{j:j \neq i}^N\left[\frac{k^{3} q_{j x x x}}{\left(q_{i}-q_{j}\right)} -\frac{3 k^{2} \left(c-k q_{j x}\right) q_{j x x}}{\left(q_{i}-q_{j}\right)^2} -\frac{2 \left(c-k q_{j x}\right)^{3}}{\left(q_{i}-q_{j}\right)^3}\right]\,.
  }
 \end{array}
\eq

The model (\ref{e01}) is integrable in the sense that it has algebro-geometric solutions and equations of motion
are represented in the  Zakharov-Shabat (or Lax or zero curvature) form:
 \beq\label{e03}
 \begin{array}{c}
  \displaystyle{
 \partial_{t}{U}(z)-k\partial_{x}{V}(z)+[{U}(z), {V}(z)]=0\,,\qquad {U}(z), {V}(z)\in\Mat\,,
  }
 \end{array}
\eq
 where $U$-$V$ pair is a pair ${U}^{\hbox{\tiny{2dCM}}}(z)$, ${V}^{\hbox{\tiny{2dCM}}}(z)$ of matrix valued functions of the fields $p_j(x)$, $q_j(x)$, $j=1,...,N$ and their derivatives. They also depend on the spectral parameter $z$, and (\ref{e03}) holds true identically in $z$ (on-shell equations of motion). Explicit expression for $U$-$V$ pair
 is as follows:
 \beq\label{e031}
 \begin{array}{c}
  \displaystyle{
{U}^{\hbox{\tiny{2dCM}}}_{i j}(z)=-\delta_{i j} \Big(p_{i}+\frac{ \alpha_{i}^2}{N z} +\frac{k \alpha_{i x}}{\alpha_{i}}\Big)+\Big(1-\delta_{i j}\Big) \alpha_{j}^2  \Big(\frac{1}{q_{i}-q_{j}}- \frac{1}{N z}\Big)\,,
  }
 \end{array}
\eq
 \beq\label{e032}
 \begin{array}{c}
  \displaystyle{
{V}^{\hbox{\tiny{2dCM}}}_{i j}(z)= \delta_{i j}\left[-\frac{q_{i t}}{N z}- \frac{c \alpha_{i}^{2}}{N z^2}+\widetilde{m}_{i}^{0} -\frac{\alpha_{i t}}{\alpha_{i}}\right]+
}
 \\ \ \\
   \displaystyle{
+\left(1-\delta_{i j}\right) \alpha_{j}^{2}\left[ \frac{c}{z} \left(\frac{1}{q_{i}-q_{j}}- \frac{1}{N z}\right) -N c \left(\frac{1}{q_{i}-q_{j}}\right)^2-\widetilde{m}_{i j} \left(\frac{1}{q_{i}-q_{j}}- \frac{1}{N z}\right)\right]\,,
  }
 \end{array}
\eq
where
 \beq\label{e036}
 \begin{array}{c}
  \displaystyle{
\alpha_{i}^{2}=k q_{i x}-c\,,\quad i=1,...,N
  }
 \end{array}
\eq
and
$$
 \begin{array}{c}
  \displaystyle{
\tilde{m}_{i}^{0}=p_{i}^{2}+\frac{k^2 \alpha_{i x x}}{\alpha_{i}}+2 \kappa p_{i}
-\!\sum_{j:j \neq i}^N\left[\frac{2 \alpha_{j}^{4}+\alpha_{i}^{2}\alpha_{j}^{2}}{(q_{i}-q_{j})^2}+ \frac{4 k \alpha_{j} \alpha_{j x}}{q_{i}-q_{j}}\right],
\ \ \
\kappa=-\frac{1}{N c} \sum_{l=1}^{N} p_{l}\left(c-k q_{l x}\right),
  }
 \end{array}
$$
 \beq\label{e034}
 \begin{array}{c}
  \displaystyle{
\widetilde{m}_{i j}=p_{i}+p_{j}+2 \kappa+\frac{k \alpha_{i x}}{\alpha_{i}}-\frac{k \alpha_{j x}}{\alpha_{j}}
-\sum_{k:k \neq i, j}^N \alpha_{k}^{2} \left(\frac{1}{q_{i}-q_{k}}+\frac{1}{q_{k}-q_{j}}-\frac{1}{q_{i}-q_{j}}\right)\,,
  }
 \end{array}
\eq
In what follows we assume the center of mass frame
%\footnote{Notice that $\sum_k {\dot q}_k=0$ but $\sum_k {\dot p}_k=0$ holds true in 0+1 mechanical limit only.}
:
 \beq\label{e037}
 \begin{array}{c}
  \displaystyle{
\sum\limits_{k=1}^N q_k=0\,.
  }
 \end{array}
\eq
Notice that in our previous paper on this topic \cite{AtZ1} we used slightly different normalization coefficients and the
gauge choice for $U$-$V$ pair, which was more convenient for the case $N=2$ when $q_1=-q_2$.
\paragraph{Limit to 0+1 mechanics.} The finite-dimensional classical mechanics appears in the limit $k\rightarrow 0$. All the fields become independent of $x$, and the field Poisson brackets turn into the ordinary Poisson brackets for mechanical $N$-body system:
 \beq\label{e04}
 \begin{array}{c}
  \displaystyle{
\{q_i,p_j\}=\delta_{ij}\,,\qquad \{p_i,p_j\}=\{q_i,q_j\}=0\,.
  }
 \end{array}
\eq
  The Hamiltonian density (\ref{e01}) in this limit provides the ordinary Calogero-Moser model \cite{Calogero, Moser}:
 \beq\label{e05}
 \begin{array}{c}
  \displaystyle{
H^{\hbox{\tiny{2dCM}}}|_{k=0}=2c H^{\hbox{\tiny{CM}}}-\frac{c}{N}\left(\sum_{i=1}^{N} {p}_{i}\right)^{2}=2c H^{\hbox{\tiny{CM}}}\,,\qquad
 H^{\hbox{\tiny{CM}}}=\sum\limits_{k=1}^N\frac{p_k^2}{2}
 -\frac12\sum\limits_{i\neq j}^N\frac{c^2}{(q_i-q_j)^2}\,,
  }
 \end{array}
\eq
 where $|_{k=0}$ in the l.h.s. assumes also transition to $x$-independent variables. Similarly,
 the Zakharov-Shabat equation (\ref{e03}) reduces to the Lax equation:
 \beq\label{e06}
 \begin{array}{c}
  \displaystyle{
\partial_{t}L^{\hbox{\tiny{CM}}}(z)+[L^{\hbox{\tiny{CM}}}(z), M^{\hbox{\tiny{CM}}}(z)]=0\,,
\qquad L^{\hbox{\tiny{CM}}}(z), M^{\hbox{\tiny{CM}}}(z)\in\Mat
}
\\ \ \\
  \displaystyle{
L_{ij}^{\hbox{\tiny{CM}}}(z)=U_{ij}^{\hbox{\tiny{2dCM}}}(z)|_{k=0}
=\delta_{i j} \left(-p_{i} +\frac{c}{N z}\right)-\left(1-\delta_{i j}\right) c  \left(\frac{1}{q_{i}-q_{j}}- \frac{1}{N z}\right)\,,
}
\\ \ \\
  \displaystyle{
M^{\hbox{\tiny{CM}}}(z)=V^{\hbox{\tiny{2dCM}}}(z)|_{k=0}=\Big(L^{\hbox{\tiny{CM}}}(z)\Big)^2+M'(z)\,,
}
\\
  \displaystyle{
  M'_{ij}(z)=-\delta_{i j} \sum_{k:k\neq i}^N \frac{2 c^2}{(q_{i}-q_{k})^2}+\left(1-\delta_{i j}\right) \frac{2 c^2}{(q_{i}-q_{j})^2}\,.
  }
 \end{array}
\eq

 \paragraph{Purpose of the paper.} The 1+1 field generalizations under consideration are widely known for the Toda chains \cite{MOP}. For the relativistic models of Ruijsenaars-Schneider type the field generalizations were proposed recently in \cite{ZZ}. In \cite{LOZ} the results of
 \cite{Krich2, Akhmetshin} were extended to (multi)spin generalizations of the Calogero-Moser model. It was also explained (using modification of bundles and the symplectic Hecke correspondence) that the field Calogero-Moser system should be gauge equivalent to some model of Landau-Lifshitz type. That is, there exist a gauge transformation $G(z)\in\Mat$, which transforms
 $U$-$V$ pair for the field Calogero-Moser model to the one for some Landau-Lifshitz type model:
 \beq\label{e07}
 \begin{array}{c}
  \displaystyle{
 U^{\hbox{\tiny{LL}}}(z)=G(z)U^{\hbox{\tiny{2dCM}}}(z)G^{-1}(z)+k\p_x G(z)G^{-1}(z)\,.
  }
 \end{array}
\eq
 For the $N=2$ case explicit construction of the matrix $G(z)$ and the change of variables  was derived in our paper \cite{AtZ1}, and the Landau-Lifshitz model for ${\rm GL}_2$ rational $R$-matrix was derived in \cite{LOZ142}.
 The goal of this article is to define the gauge transformation in ${\rm gl}_N$ case, describe the corresponding Landau-Lifshitz type model and find explicit change of variables using relation  (\ref{e07}).

%%%%%%%%%%%%%%%%%%%%%%%%%%%%%%%%%%%%%%%%%%%%%%%%%%%%%%%%%%%%%%%%%%%%%%%%%%%%%%%%%%%%%%%%%%%%%%%%%%%%%%
%%%%%%%%%%%%%%%%%%%%%%%%%%%%%%%%%%%%%%%%%%%%%%%%%%%%%%%%%%%%%%%%%%%%%%%%%%%%%%%%%%%%%%%%%%%%%%%%%%%%%%
\section{Rational top and Landau-Lifshitz equation}
\setcounter{equation}{0}

 \paragraph{Rational integrable top.} In order to explain what kind of Landau-Lifshitz model is expected in (\ref{e07}) we first consider its 0+1 mechanical analogue.
 The mechanical version of (\ref{e07}) is as follows:
 \beq\label{e08}
 \begin{array}{c}
  \displaystyle{
 L^{\hbox{\tiny{top}}}(z)=g(z)L^{\hbox{\tiny{CM}}}(z)g^{-1}(z)\,,
  }
 \end{array}
\eq
 where $L^{\hbox{\tiny{top}}}(z)$ is the Lax matrix of some integrable top like model.
 It is the model, which was introduced in \cite{AASZ} and called the rational top. Equations of
 motion for top like models are of the form:
 \beq\label{e09}
 \begin{array}{c}
  \displaystyle{
 \partial_{t} S =\{S,H^{\hbox{\tiny{top}}}\}= 2 c [S, J(S)]\,,\qquad S=\sum\limits_{i,j=1}^N E_{ij}S_{ij}\in\Mat\,,
  }
 \end{array}
\eq
where $S$ is a matrix of dynamical variables ($E_{ij}$ is the standard matrix basis), $c\in\mC$ is a constant and $J(S)$ is some special linear map (see \cite{AASZ}).
The Hamiltonian is quadratic, and the  Poisson brackets are given by the Poisson-Lie structure on ${\rm gl}^*_N$ Lie coalgebra:
 \beq\label{e10}
 \begin{array}{c}
  \displaystyle{
 H^{\hbox{\tiny{top}}}=cN  \tr\Big(SJ(S)\Big)\,,\qquad
 \left\{S_{i j}, S_{k l}\right\}=\frac{1}{N}\,\Big(S_{i l} \delta_{k j}-S_{k j} \delta_{i l}\Big)\,.
  }
 \end{array}
\eq
It was shown in \cite{AASZ} that in the special
case ${\rm rk}(S)=1$ (and $\tr(S)=c$) this model is gauge equivalent (\ref{e08}) to the rational Calogero-Moser model. Namely, it was proved by direct evaluation that the expression in the r.h.s. of (\ref{e08}) is represented in the form:
 \beq\label{e11}
 \begin{array}{c}
  \displaystyle{
 g(z)L^{\hbox{\tiny{CM}}}(z)g^{-1}(z)=\tr_2\Big(r_{12}(z)\stackrel{2}{S}\Big)\,,\quad \stackrel{2}{S}=1_N\otimes S\,,
  }
 \end{array}
\eq
where $S_{ij}=S_{ij}(p_1,...,p_N,q_1,...,q_N,c)$, $r_{12}(z)$ is some classical non-dynamical $r$-matrix (satisfying the classical Yang-Baxter equation),
 $1_N$ is the identity $N\times N$ matrix and $\tr_2$ means trace over the second tensor component in $\Mat^{\otimes 2}$. The gauge equivalence means that the Hamiltonians $H^{\hbox{\tiny{top}}}$ (\ref{e10}) and
 $H^{\hbox{\tiny{CM}}}$ (\ref{e05}) coincide under a certain change of variables, which will be given below in (\ref{e86}).

 \paragraph{Description through $R$-matrix.} In \cite{LOZ16} a construction of Lax pairs with spectral parameter was suggested based on (skew-symmetric and unitary) solution of the associative
Yang-Baxter equation \cite{FK, Polishchuk}:
 \beq\label{e12}
 \begin{array}{c}
  \displaystyle{
 R^{\hbar}_{12} R^{\eta}_{23} = R^{\eta}_{13} R^{\hbar-\eta}_{12} + R^{\eta-\hbar}_{23} R^{\hbar}_{13}\,,
   \qquad R^x_{ab} = R^x_{ab}(z_a-z_b)\,.
  }
 \end{array}
\eq
In fact, a skew-symmetric and unitary solution of (\ref{e12}) in the fundamental representation of ${\rm GL}_N$ Lie group
is a quantum $R$-matrix, i.e. it satisfies also the quantum Yang-Baxter equation $R_{12}^\hbar R_{13}^\hbar R_{23}^\hbar
=R_{23}^\hbar R_{13}^\hbar R_{12}^\hbar$. Consider the classical limit expansion of such $R$-matrix:
 \beq\label{e13}
 \begin{array}{c}
  \displaystyle{
 R^{\hbar}_{12} (z)=\frac{1}{\hbar}\, 1_N \otimes 1_N + r_{12}(z) + \hbar \: m_{12} (z) + O(\hbar^2)\,.
  }
 \end{array}
\eq
Then the Lax pair can be written as follows:
 \beq\label{e14}
 \begin{array}{c}
  \displaystyle{
 L^{\hbox{\tiny{top}}}(z)=\,\tr_2\Big(r_{12}(z)\stackrel{2}{S}\Big)\,,\qquad
 M^{\hbox{\tiny{top}}}(z)=-\,\tr_2\Big(m_{12}(z)\stackrel{2}{S}\Big)\,.
  }
 \end{array}
\eq
It generates the Euler-Arnold equation (\ref{e09}) with
 \beq\label{e15}
 \begin{array}{c}
  \displaystyle{
J(S)=\tr_2\Big(m_{12}(0)\stackrel{2}{S}\Big)\,.
  }
 \end{array}
\eq

\paragraph{Rational $R$-matrix.}

In this paper we will use the rational $R$-matrix calculated in \cite{LOZ14}. In the $N=2$ case it reproduces the 11-vertex $R$-matrix
found by I. Cherednik \cite{Chered}:
 \beq\label{e17}
 \begin{array}{c}
  \displaystyle{
 R_{12}^\hbar(z)=
  \left(
  \begin{array}{cccc}
  1/\hbar+1/z & 0 & 0 & 0
  \\
  -z-\hbar & 1/\hbar & 1/z & 0
  \\
   -z-\hbar & 1/z & 1/\hbar & 0
   \\
   -z^3-\hbar^3-2z^2\hbar-2z\hbar^2 & z+\hbar & z+\hbar & 1/\hbar+1/z
   \end{array}
   \right)\,.
  }
 \end{array}
\eq
For $N>2$ all its properties, different possible forms and explicit expressions for the coefficients
of expansions (\ref{e13}) and (\ref{e1602}) can be found in \cite{AtZ3}.

\paragraph{Rational IRF-Vertex transformation.}
Following \cite{AASZ} introduce the matrix $g(z)\in\Mat$:
 \beq\label{e161}
 \begin{array}{c}
  \displaystyle{
 g(z)=g(z,q_1,...,q_N)=\Xi(z,q)D^{-1}(q)\,,\qquad \Xi(z,q)\,,D(q)\in\Mat
  }
 \end{array}
\eq
where
   \begin{equation}
   \label{e162}
   \begin{array}{c}
   D_{ij}({q}) =\delta_{ij}\prod\limits_{k \neq i}^{N}(q_{i}-q_{k})\,,
     \qquad
   \Xi_{ij}(z,{q}) =(z+q_{j})^{\varrho(i)}\,,
   \qquad
   \sum\limits_{k=1}^{N}q_{k}=0\,,
   \end{array}
   \end{equation}
   with
\begin{equation}
   \label{e163}
   \varrho(i)=\left\{\begin{array}{ll}
i-1 \quad \text { for } 1 \leq i \leq N-1\,,  \\
i \quad \text { for } i=N\,,
\end{array}\right. ~~~~\varrho^{-1}(i)=\left\{\begin{array}{cl}
i+1 & \text { for } 0 \leq i \leq N-2\,, \\
i & \text { for } i=N\,.
\end{array}\right.
   \end{equation}
The matrix $\Xi(z)$ is degenerated at $z=0$: $\det\Xi(z,q)=Nz\prod\limits_{i>j}^N(q_i-q_j)$.
It plays the role of IRF-Vertex transformation for rational $R$-matrices \cite{AtZ3}. The inverse
of matrix $g(z,q)$ is as follows:
   \beq\label{e84}
 \begin{array}{c}
 \displaystyle{
g^{-1}_{kj}({z,q})=(-1)^{\varrho(j)}\Big(\frac{\sigma_{\varrho(j)}({x})}{Nz}\,-
\stackrel{k}{\sigma}_{\varrho(j)}({x})\Big)\,,\quad  x_j=z+q_j\,,
%x_j=z+q_j-\frac1N\sum\limits_{k=1}^N q_k\,.
}
 \end{array}
 \eq
where $\sigma_j(x)$ and $\stackrel{k}{\sigma}_j(x)$ are symmetric functions (for variables $x_1,...x_N$) defined as
 \beq\label{e85}
  \begin{array}{c}
  \displaystyle{
%\sigma_{N-d}(\mathbf{x})=(-1)^{N} \sum_{1 \leq i_{1}<i_{2} \ldots<i_{d} \leq N} x_{i_{1}} x_{i_{2}} \ldots x_{i_{d}}, %\quad d=0, \ldots, N
\prod\limits_{m=1}^{N} \,(\zeta-x_{m})=
\sum\limits_{k=0}^{N} (-1)^{k} \zeta^{k} \sigma_{k}(x_{1},...,x_{N})\,,
\qquad
\prod\limits_{m:m\neq k}^{N} \,(\zeta-x_{m})=-\sum\limits_{s=0}^{N-1} (-1)^{s} \zeta^{s}
\stackrel{k}{\sigma}_s(x)\,.
 }
 \end{array}
 \eq
Details can be found in \cite{AASZ,AtZ3}.
The latter formula provides via (\ref{e08}), (\ref{e11}) explicit change of variables in 0+1 mechanics between
Calogero-Moser model (\ref{e06}) and the rational top (\ref{e09})-(\ref{e10}), (\ref{e14})-(\ref{e15}):
 \beq\label{e86}
  \begin{array}{c}
  \displaystyle{
S_{i j}= \frac{(-1)^{\varrho(j)}}{N} \sum_{m=1}^{N} \frac{-(q_{m})^{\varrho(i)} \check{p}_{m} +c \varrho(i) (q_{m})^{\varrho(i)-1}}{\prod_{l \neq m}\left(q_{m}-q_{l}\right)}\,  \sigma_{\varrho(j)}(q)\,,
\quad
\check{p}_j=p_j+\sum\limits_{l:l\neq j}\frac{c}{q_j-q_l}\,.
 }
 \end{array}
 \eq
 Similar results are known for trigonometric \cite{KrZ} and elliptic \cite{LOZ,ZZ} models.

\paragraph{Landau-Lifshitz equation.}
Recently the 1+1 field generalization of the Lax pair (\ref{e14}) to $U$-$V$ pair was suggested in  \cite{AtZ2}.
In the field case the Poisson brackets (\ref{e10}) are replaced with
 \beq\label{e16}
 \begin{array}{c}
  \displaystyle{
 \left\{S_{i j}(x), S_{k l}(y)\right\}=\frac{1}{N}\,\Big(S_{i l}(x) \delta_{k j}-S_{k j}(x) \delta_{i l}\Big)\delta(x-y)\,.
  }
 \end{array}
\eq
The construction of $U$-$V$ pair is again based on $R$-matrix satisfying the associative Yang-Baxter equation (\ref{e12}). For this purpose the following relation is used (it can be deduced from (\ref{e12})):
 \beq\label{e1601}
  \begin{array}{c}
  \displaystyle{
  r_{12}(z)r_{13}(z)=r_{23}^{(0)}r_{12}(z)-r_{13}(z)r_{23}^{(0)}
  -\p_z r_{13}(z)P_{23}+m_{12}(z)+m_{23}(0)+m_{13}(z)\,,
 }
 \end{array}
 \eq
 where $P_{12}$ is the matrix permutation operator and $r_{12}^{(0)}$ is the coefficient in the expansion
 \beq\label{e1602}
  \begin{array}{c}
  \displaystyle{
 r_{12}(z)=z^{-1}P_{12}+r_{12}^{(0)}+O(z)\,.
 }
 \end{array}
 \eq
 Suppose ${\rm rank}(S)=1$, so that $S^2=cS$, $c=\tr(S)$. Then the Landau-Lifshitz equation reads:
 \beq\label{q51}
  \begin{array}{c}
  \displaystyle{
  \p_t S=\frac{k^{2}}{c}\,[S,\p^2_x S]+ 2c \, [S,J(S)]-2 k [S,E(\p_x S)]\,,
 }
 \end{array}
 \eq
where
 \beq\label{q33}
  \begin{array}{c}
  \displaystyle{
 E(S)=\,\tr_2\Big(r^{(0)}_{12}\stackrel{2}{S}\Big)\,,\quad \stackrel{2}{S}=1_N\otimes S\,,\quad  S\in\Mat\,.
 }
 \end{array}
 \eq
Then the $U$-$V$ pair generating equations of motion (\ref{q51}) through the Zakharov-Shabat equation (\ref{e03})
has the form:
 \beq\label{q34}
  \begin{array}{c}
  \displaystyle{
 U^{\hbox{\tiny{LL}}}(z)=L^{\hbox{\tiny{top}}}(S,z)=\,\tr_2\Big(r_{12}(z)\stackrel{2}{S}\Big)\,,\qquad
 V^{\hbox{\tiny{LL}}}(z)=V_1(z)+V_2(z)\,,
 }
 \end{array}
 \eq
 \beq\label{q35}
  \begin{array}{c}
  \displaystyle{
 V_1(z)=-c\p_z L^{\hbox{\tiny{top}}}(S,z)+L^{\hbox{\tiny{top}}}(E(S)S,z)\,,\quad
 V_2(z)=-cL^{\hbox{\tiny{top}}}(T,z)\,,\quad T=-\frac{k}{c^2}\,[S,\p_x S]\,.
 }
 \end{array}
 \eq
Equations (\ref{q51}) are Hamiltonian with the following Hamiltonian function:
 \beq\label{q53}
  \begin{array}{c}
  \displaystyle{
 H^{\hbox{\tiny{LL}}}=\oint dy\Big( cN  \tr\Big(S\,J(S)\Big)-\frac{N k^{2}}{2c}\,\tr\Big(\p_yS\,\p_yS\Big)+
 kN  \tr\Big(\p_y S\,E(S)\Big)\Big)\,,\qquad S=S(y)\,,
 }
 \end{array}
 \eq
so that (\ref{q51}) is reproduced as $\p_tS(x)=\{S(x),H^{\hbox{\tiny{LL}}}\}$ with the Poisson brackets (\ref{e16}).

%%%%%%%%%%%%%%%%%%%%%%%%%%%%%%%%%%%%%%%%%%%%%%%%%%%%%%%%%%%%%%%%%%%%%%%%%%%%%%%%%%%%%%%%%%%%%%%%%%%%%%
%%%%%%%%%%%%%%%%%%%%%%%%%%%%%%%%%%%%%%%%%%%%%%%%%%%%%%%%%%%%%%%%%%%%%%%%%%%%%%%%%%%%%%%%%%%%%%%%%%%%%%
\section{Gauge equivalence and change of variables}
\setcounter{equation}{0}

Introduce the matrix $G(z,q)=b(x,t) g(z,q)$, where $b(x,t)$ is the following function:
 \beq\label{e81}
  \begin{array}{c}
  \displaystyle{
G(z, q)=b(x, t) \Xi(z, {q})D^{-1}\in\Mat\,,
\quad
b(x,t)=\prod _{a<b}^N (q_{b}-q_{a})^{1/N} \prod _{m=1}^N\left(k q_{m,x}-c \right)^{1/(2N)}\,.
 }
 \end{array}
 \eq
The statement is that by applying the gauge transformation with the matrix (\ref{e81}) we obtain
the desired relation
 (\ref{e07})\footnote{Let us also remark that $V$-matrices of 1+1 Calogero-Moser and the Landau-Lifshitz models are also related by the gauge transformation $ V^{\hbox{\tiny{LL}}}(z)=G(z)V^{\hbox{\tiny{2dCM}}}(z)G^{-1}(z)+\p_t G(z)G^{-1}(z)$ up
to additional scalar (i.e. proportional to $1_N$) term. The latter can be removed by applying
additional gauge transformation with the matrix $G=\exp\left(-\frac{c^2 t}{N z^2}-\int_{t_0}^t f(x, t') dt' \right) 1_N  $, where $f(x, t)=\frac{1}{N}\sum_{i=1}^N \tilde{m}_{i}^{0} $. 
%This transformation does not change $U(z)$.
}.
 Calculations are performed similarly to those in 0+1 mechanics \cite{AASZ}.
As a result we obtain explicit change of variables:
 \beq\label{e88}
  \begin{array}{c}
  \displaystyle{
S_{i j}= \frac{(-1)^{\varrho(j)+1}}{N} \sum_{m=1}^{N}  \frac{(q_{m})^{\varrho(i)} (\tilde{p}_{m} +\frac{k \alpha_{m x}}{\alpha_{m}})+\alpha_{m}^{2} \varrho(i) (q_{m})^{\varrho(i)-1}}{\prod_{l \neq m}\left(q_{m}-q_{l}\right)} \,\sigma_{\varrho(j)}({q})\,,\ \
\tilde{p}_{j}=p_{j}- \sum_{l \neq j}^{N} \frac{\alpha_{j}^{2}}{q_{j}-q_{l}}
 }
 \end{array}
 \eq
with the properties
 \beq\label{e89}
  \begin{array}{c}
  \displaystyle{
  Spec(S)=(0,...,0, c)\,,\qquad {\rm rk}(S)=1\,,\qquad
  \tr(S)= c\,,\qquad S^2=  c S\,.
   }
 \end{array}
 \eq
It is the 1+1 field generalization of the change of variables in mechanics (\ref{e86}). It can be also verified that
the Poisson brackets for $S_{i j}(p,q,c)$ (\ref{e88}) calculated through the canonical brackets (\ref{e02})
indeed reproduce the linear Poisson structure (\ref{e16}), so that (\ref{e88}) is a Poisson map. The Hamiltonian (\ref{e01}) of 1+1 field Calogero-Moser model coincides with the one (\ref{q53}) for the Landau-Lifshitz equation under the change of variables (\ref{e88}): $H^{\hbox{\tiny{LL}}}[S(p(x),q(x))]=H^{\hbox{\tiny{2dCM}}}[p(x),q(x)]$.

\paragraph{Acknowledgments.}
This work was supported by the Russian Science Foundation under grant number 21-41-09011, https://rscf.ru/en/project/21-41-09011/.

%%%%%%%%%%%%%%%%%%%%%%%%%%%%%%%%%%%%%%%%%%%%%%%%%%%%%%%%%%%%%%%%%%%%%%%%%%%%%%%%%%%%%%%%%%%%%%%%%%%%%%
%%%%%%%%%%%%%%%%%%%%%%%%%%%%%%%%%%%%%%%%%%%%%%%%%%%%%%%%%%%%%%%%%%%%%%%%%%%%%%%%%%%%%%%%%%%%%%%%%%%%%%
%\section{Conclusion}
%\setcounter{equation}{0}

%%%%%%%%%%%%%%%%%%%%%%%%%%%%%%%%%%%%%%%%%%%%%%%%%%%%%%%%%%%%%%%%%%%%%%%%%%%%%%%%%%%%%%%%%%%%%%
%%%%%%%%%%%%%%%%%%%%%%%%%%%%%%%%%%%%%%%%%%%%%%%%%%%%%%%%%%%%%%%%%%%%%%%%%%%%%%%%%%%%%%%%%%%%%%

\begin{small}

\end{small}

\end{document}